\documentclass[
reprint,
superscriptaddress,
amsmath,
amssymb,
aps,
longbibliography,
pra, 
showpacs,
floatfix
]{revtex4-1}

\usepackage{graphicx}
\usepackage{color}

\usepackage{tabularx}
\usepackage{multirow}
\usepackage{dcolumn} 
\newcolumntype{d}[1]{D{.}{.}{#1}}
\newcolumntype{L}[1]{>{\raggedright\arraybackslash}p{#1}}
\newcolumntype{C}[1]{>{\centering\arraybackslash}p{#1}}
\newcolumntype{G}[1]{>{\raggedleft\arraybackslash}p{#1}}
\usepackage{booktabs}

\usepackage{comment}
\usepackage{enumitem}

\newcommand{\ang}{\ensuremath{\mathrm{\AA}}}

\begin{document}

\title{Parametric excitation of an optically silent Goldstone-like phonon mode}

\author{Dominik~M.\ Juraschek}
\email{djuraschek@seas.harvard.edu}
\affiliation{Harvard John A. Paulson School of Engineering and Applied Sciences, Harvard University, Cambridge, MA 02138, USA}
\author{Quintin N. Meier}
\affiliation{Department of Materials, ETH Zurich, CH-8093 Z\"urich, Switzerland}
\author{Prineha Narang}
\email{prineha@seas.harvard.edu}
\affiliation{Harvard John A. Paulson School of Engineering and Applied Sciences, Harvard University, Cambridge, MA 02138, USA}
\date{\today}

\begin{abstract}
It has recently been indicated that the hexagonal manganites exhibit Higgs- and Goldstone-like phonon modes that modulate the amplitude and phase of their primary order parameter. Here, we describe a mechanism by which a silent Goldstone-like phonon mode can be coherently excited, which is based on nonlinear coupling to an infrared-active Higgs-like phonon mode. Using a combination of first-principles calculations and phenomenological modeling, we describe the coupled Higgs-Goldstone dynamics in response to the excitation with a terahertz pulse. Besides theoretically demonstrating coherent control of crystallographic Higgs and Goldstone excitations, we show that the previously inaccessible silent phonon modes can be excited coherently with this mechanism.
\end{abstract}

\maketitle



Order parameters are physical observables that are used to quantify the different states of matter. Their amplitudes and phases can be excited by external stimuli, such as a laser pulse, leading to exotic states of matter that cannot be accessed in equilibrium \cite{Basov2017}. Two particular excitations are Higgs and Goldstone modes, which correspond to the modulation of the amplitude and phase of an order parameter that breaks a continuous symmetry. Originally discussed in the context of particle physics \cite{Higgs2014} and superconductivity \cite{Anderson2015}, Higgs and Goldstone excitations were found in cold atom systems, such as superfluids \cite{Podolsky2011,Pollet2012,Endres2012,Hoinka2017,Behrle2018} or supersolids \cite{Leonard2017}. Higgs and Goldstone modes are well-studied in superconductors today \cite{Varma2002,Barlas2013,Matsunaga2013,Measson2014,Matsunaga2014,Sherman2015,Kemper2015,Pekker2015,Fauseweh2017,Chu2019,Shimano2019,Buzzi2019}, and similar manifestations have recently been reported in charge density wave (CDW) systems \cite{Liu2013,Mankowsky2017_3,Zong2018}, antiferromagnets \cite{Ruegg2008,Jain2017,Souliou2017}, and excitonic insulators \cite{Werdehausen2018}. It has been debated whether the optical and acoustic vibrational modes of solids can be considered Higgs and Goldstone excitations of the crystal lattice \cite{Vallone2019}. Several complex transition metal oxide compounds have shown signatures of \textit{optical} Goldstone-like phonon modes that live in their symmetry-broken potential energy landscape \cite{Kendziora2005,Nakhmanson2010,Mangeri2016,Marthinsen2018,Meier2019}.

Coherent control over Raman-active phonons via impulsive stimulated Raman scattering using visible light pulses is well-established \cite{desilvestri:1985,merlin:1997}. Recently, the excitation of infrared (IR)-active phonons with large amplitudes via IR absorption has become feasible through the development of high intensity terahertz and mid-IR sources. This progress has enabled selective control over the dynamics of the crystal lattice through nonlinear phonon interactions using ionic Raman scattering \cite{forst:2011,subedi:2014,nicoletti:2016} and two-particle absorption mechanisms \cite{maehrlein:2017, Juraschek2018,Melnikov2018,Kozina2019,Johnson2019,Knighton2019}. In contrast, phonon modes that do not respond to IR absorption or Raman scattering techniques, so called silent modes, can only be detected through hyper-Raman scattering. As hyper-Raman scattering is a third-order interaction of the electric field component of light with a phonon mode, it is inefficient and no coherent excitation has been achieved yet.

\begin{figure}[b]
\centering
\includegraphics[scale=0.06]{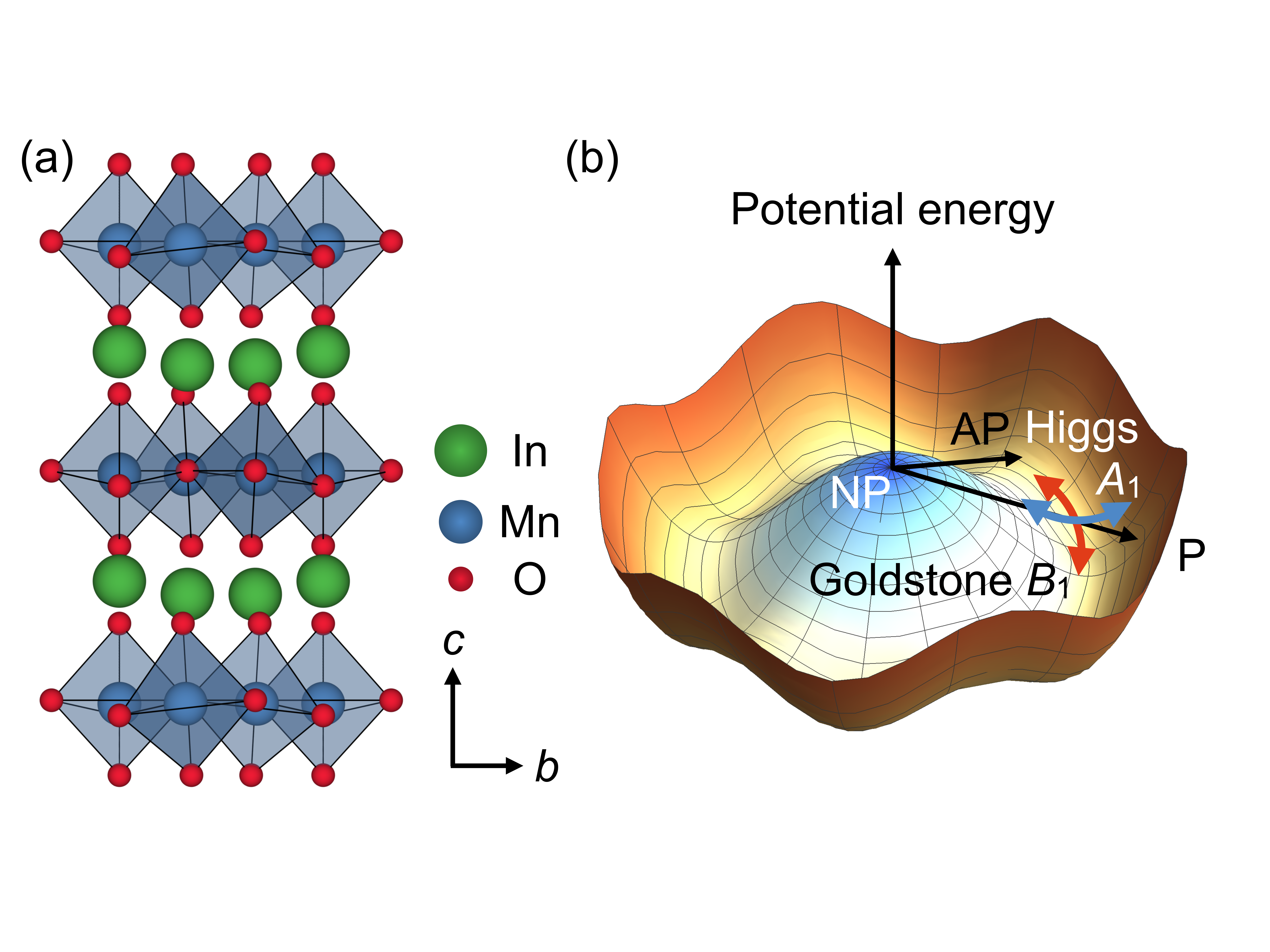}
\caption{Crystallographic properties of InMnO$_3$. (a) View of the $bc$ plane of the crystal in its hexagonal phase with space group $P6_3cm$. (b) Mexican hat potential energy landscape of the lattice with nonpolar (NP), antipolar (AP), and polar (P) phases indicated. Blue and red arrows depict the Higgs-like $A_1$ and Goldstone-like $B_1$ modes in the polar phase.
}
\label{fig:inmno3}
\end{figure}

\begin{figure}[t]
\centering
\includegraphics[scale=0.075]{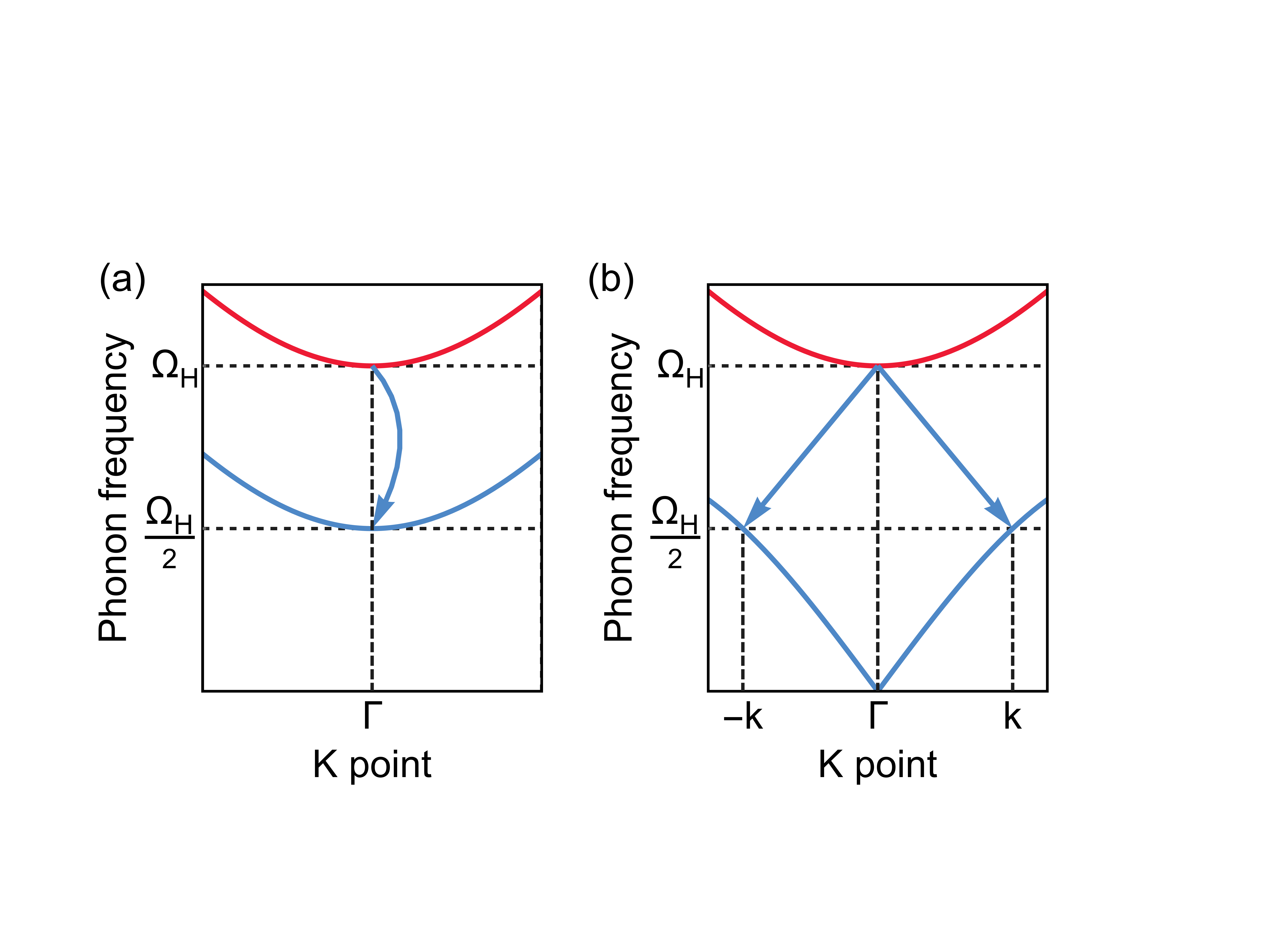}
\caption{Schematic parametric downconversion mechanisms of the resonantly driven Higgs mode (red). (a) Downconversion to the Goldstone mode (blue) at the Brillouin zone center, as proposed in this work. (b) Conventional parametric decay into acoustic modes (blue) with opposite wave vectors throughout the Brillouin zone.
}
\label{fig:downconversion}
\end{figure}

Higgs and Goldstone modes, similar to silent phonon modes, are challenging to excite and coherently address, because they often do not directly couple to light. In this work, we overcome this hurdle by synthesizing advances from two parallel fields, order-parameter physics of continuous symmetry broken systems and coherent control of crystal lattice dynamics. Specifically, we demonstrate that the low-frequency silent Goldstone-like optical phonon mode in indium manganite (InMnO$_3$) can be coherently excited by coupling nonlinearly to the low-frequency Higgs-like optical phonon mode. In InMnO$_3$, this Higgs-like phonon mode is IR active and can be driven resonantly with a terahertz pulse. We depict the two phonon modes in Fig.~\ref{fig:inmno3}.
The underlying mechanism of the excitation of the Goldstone-like phonon mode is a parametric downconversion process from an optical phonon mode to another optical phonon mode at the Brillouin zone center as shown in Fig.~\ref{fig:downconversion}(a), enabled through the nonlinear phonon coupling. This process results in a coherent excitation of the silent Goldstone-like phonon mode with large amplitude. The mechanism of parametric downconversion has previously been limited to the conventional decay of gapped (or optical) excitations into gapless (or acoustic) excitations with high wave vectors as shown in Fig.~\ref{fig:downconversion}(b), where experimental demonstrations include phonons in bismuth \cite{Teitelbaum2018}, magnons in calcium ruthenate \cite{Jain2017}, and charge density wave modes in blue bronze \cite{Liu2013}.


\paragraph*{Structural properties of InMnO$_3$.}

InMnO$_3$ crystallizes in the hexagonal manganite structure shown in Fig.~\ref{fig:inmno3}(a): The potential energy landscape of the lattice displacement forms a buckled Mexican hat that hosts a nonpolar phase at its center, and polar and antipolar phases at the minima and maxima of its brim, see Fig.~\ref{fig:inmno3}(b) \cite{Griffin2012,Artyukhin2014,Skjaervo2019}. The primary order parameter corresponds to a tilting of the manganese-oxygen bipyramids and a simultaneous buckling of the indium atoms. This two-dimensional order parameter couples to a ferroelectric displacement with polarization along the $c$-axis of the crystal. This coupling is responsible for the minima in the brim of the Mexican hat and stabilizes the improper ferroelectric ground state \cite{Artyukhin2014}. InMnO$_3$ is a special member of the hexagonal manganite family and has a particularly flat brim which can be tuned using chemical doping or strain \cite{kumagaiObservationPersistentCentrosymmetricity2012,Huang_et_al:2014}. The 30 atom primitive unit cell of the polar phase with \textit{P6$_3$cm} space group (\textit{6mm} point group) hosts 90 phonon modes with irreducible representations $A_{1,2}$, $B_{1,2}$, and $E_{1,2}$. It has recently been indicated that the low-frequency $A_1$ mode at 4~THz and $B_1$ mode at 2.4~THz correspond to the amplitude and phase modulations of the primary order parameter and can therefore be considered to have Higgs and Goldstone character \cite{Meier2019}. The Higgs-like $A_1$ mode is Raman-active and IR-active along the hexagonal crystal axis, while the Goldstone-like $B_1$ mode is silent (neither Raman, nor IR active). For simplicity, we will in the following refer to the $A_1$ as ``Higgs'' mode and to the $B_1$ as ``Goldstone'' mode.


\paragraph*{Density functional theory calculations.}


We begin by calculating the structural properties of InMnO$_3$ from first principles using the density functional theory formalism as implemented in the \textsc{abinit} code \cite{Gonze2016}. For the exchange-correlation functional, we apply the local density approximation plus Hubbard $U$ (LDA+$U$) \cite{Amadon:2008ia} and we use the default projector augmented wave (PAW) method \cite{Torrent:2008jw} with GPAW pseudopotentials \cite{GPAW}. Good convergence was achieved using a cutoff energy of 30~Ha and a gamma-centered $k$-point mesh of 6$\times$6$\times$4. We apply a Hubbard $U$ of 4.5~eV and exchange $J$ of 0.45~eV using frustrated antiferromagnetic ordering \cite{Medvedeva_2000}. We obtain the phonon eigenfrequencies, eigenvectors, and Born effective charge tensors using density functional perturbation theory \cite{Gonze1997,Gonze1997_2}. In order to obtain the nonlinear phonon couplings we calculate the total energy as a function of ion displacements along the normal mode coordinates of the Higgs and Goldstone modes and then fit the resulting two-dimensional energy landscape to the potential $V$ in Eqn.~(\ref{eq:phononpotential}).



\begin{table}[b]
\centering
\bgroup
\def\arraystretch{1.5}
\caption{
Calculated eigenfrequencies $\Omega_\mathrm{H/G}$, coupling coefficients $c$ to $g$, and mode effective charge $Z_\mathrm{H}$, and peak electric field $E_0$ and full width at half maximum pulse duration $\tau$ of the laser pulse. The coupling coefficients are in units of meV/(\AA$\sqrt{\mathrm{amu}}$)$^n$, where $n$ is the order of phonon amplitude. 
}
\begin{tabular*}{\linewidth}{@{\extracolsep{\fill}}ccccccccccc}
\hline\hline
\multicolumn{1}{c}{$\Omega_\mathrm{H}/(2\pi)$} &
\multicolumn{1}{c}{$\Omega_\mathrm{G}/(2\pi)$} &
\multicolumn{1}{c}{$c$} &
\multicolumn{1}{c}{$d$} &
\multicolumn{1}{c}{$e$} &
\multicolumn{1}{c}{$f$} &
\multicolumn{1}{c}{$g$} &
\multicolumn{1}{c}{$Z_\mathrm{H}$} &
\multicolumn{1}{c}{$E_0$} & 
\multicolumn{1}{c}{$\tau$} \\
\hline
4.0 THz & 2.4 THz & 9.2 & 0.9 & 1.5 & 0.2 & 0.5 & 2.7 $e$ & 1 $\frac{\text{MV}}{\text{cm}}$ & 0.5 ps\\
\hline\hline
\end{tabular*}
\label{tab:inmno3}
\egroup
\end{table}


\paragraph*{Nonlinear Higgs-Goldstone dynamics.}


\begin{figure*}[t]
\centering
\includegraphics[scale=0.095]{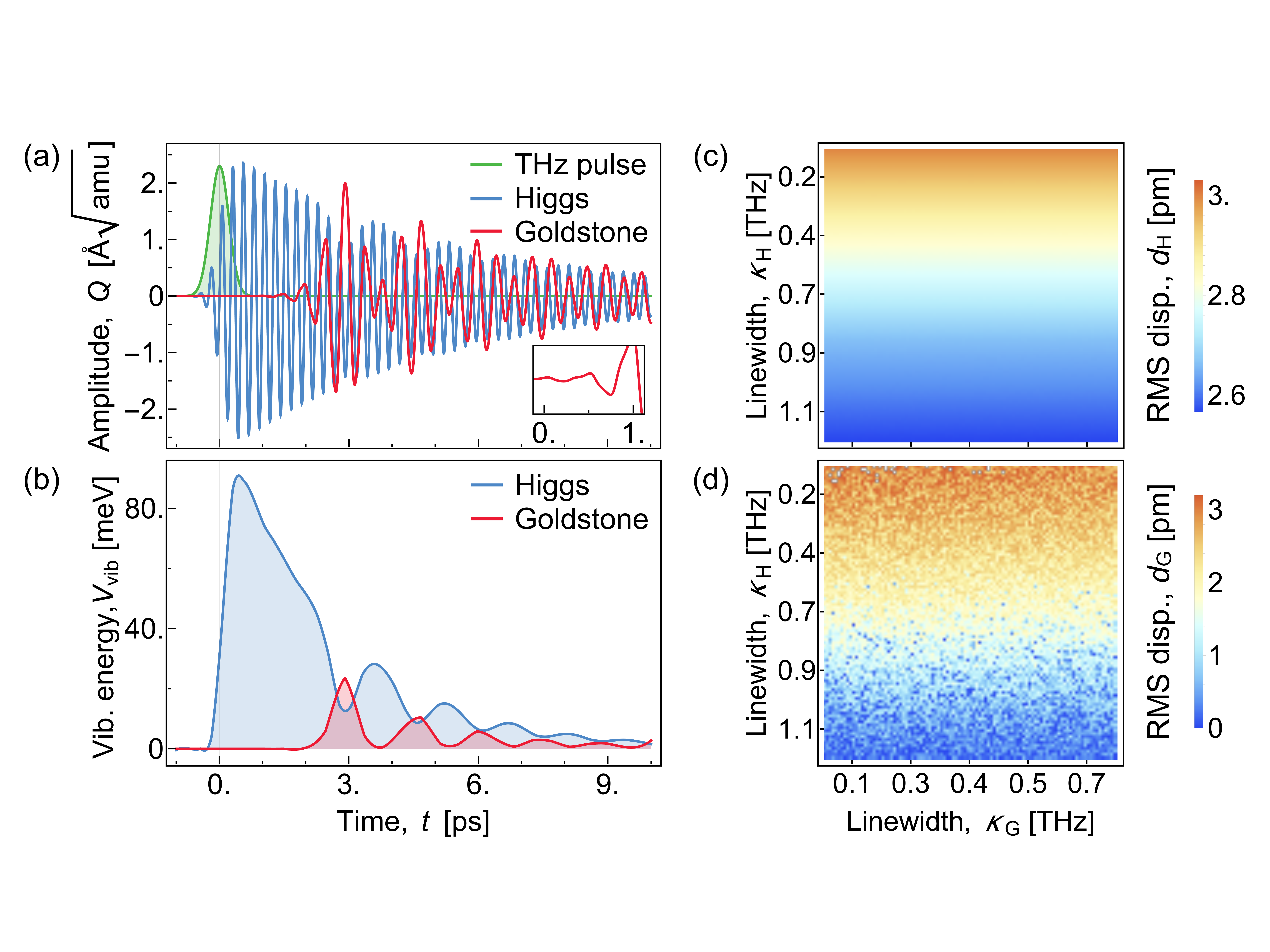}
\caption{Phonon dynamics in response to the excitation by a pulse with center frequency $\omega_0=4$~THz, duration $\tau=0.5$~ps, and peak field $E_0=1$~MV/cm. (a) Time evolution of the amplitudes $Q$ of the 4~THz Higgs and 2.4~THz Goldstone modes for linewidths of $\kappa_\mathrm{H}=0.4$~THz and $\kappa_\mathrm{G}=0.24$~THz. We show the carrier envelope of the pulse schematically. The inset shows the Goldstone mode resolved in the region between 0 and 1~ps. (b) Evolution of the vibrational energy per unit cell $V_\mathrm{vib}$ contained in each of the phonon modes in units of eV, extracted from the root mean squares (RMSs) of the phonon amplitudes. (c) and (d) RMSs of the largest ionic displacements along the Higgs ($d_\mathrm{H}$) and Goldstone ($d_\mathrm{G}$) modes for different $\kappa_\mathrm{H}$ and $\kappa_\mathrm{G}$ ranging between 3--20\%{} of their phonon frequencies.
}
\label{fig:phonondynamics}
\end{figure*}

To investigate the time evolution of the Higgs and Goldstone modes in response to a pulsed terahertz excitation, we numerically solve their equations of motion
\begin{equation}\label{eq:eom}
\ddot{Q}_i + \kappa_i Q_i + \frac{\partial V}{\partial Q_i} - Z_i E = 0,
\end{equation}
where $Q_i$ is the normal mode coordinate (or amplitude) of the Higgs ($i$=H) and Goldstone ($i$=G) mode in units of $\ang\sqrt{\mathrm{amu}}$, where amu is the atomic mass unit. $\kappa_i$ is the phonon linewidth, $Z_i$ is the mode effective charge, $V(Q)$ is the anharmonic phonon potential, and $E$ the electric field component of the pulse. The mode effective charge is given by $Z_i=\sum_n Z^\ast_n \mathbf{q}_{i,n} / \sqrt{M_n}$, where $Z^\ast_n$ is the Born effective charge tensor of atom $n$ in units of the elementary charge $e$, $\mathbf{q}_{i,n}$ is its eigenvector, and $M_n$ its atomic mass \cite{Gonze1997}. $Z_i$ is only nonzero if $i$ is an IR-active phonon mode. The displacement of each individual ion $\mathbf{d}_{i,n}$ along the phonon modes can be calculated as $\mathbf{d}_{i,n} = \mathbf{q}_{i,n} Q_i(t)/\sqrt{M_n}$.

As $A_1\subset[B_1 \times B_1]$, the lowest-order symmetry-allowed nonlinear phonon coupling between the Higgs and Goldstone modes is of the form $Q_\mathrm{H}Q_\mathrm{G}^2$. A phonon coupling squared in the IR-active mode and linear in the other one that is required for ionic Raman scattering \cite{subedi:2014,Juraschek2018} is forbidden by symmetry for the silent $B_1$ mode here. The potential energy of the phonons can therefore be written in a minimal model as
\begin{equation}\label{eq:phononpotential}
V_\mathrm{min}(Q) = \frac{\Omega_\mathrm{H}^2}{2} Q_\mathrm{H}^2 + \frac{\Omega_\mathrm{G}^2}{2} Q_\mathrm{G}^2 + c Q_\mathrm{H} Q_\mathrm{G}^2,
\end{equation}
where $\Omega_\mathrm{H}$ and $\Omega_\mathrm{G}$ are the phonon eigenfrequencies given in 2$\pi\times$THz, and $c$ is their linear-quadratic coupling coefficient given in meV/(\ang{}$\sqrt{\mathrm{amu}}$)$^3$. For accuracy we include all anharmonicities and nonlinear phonon couplings up to fourth order in the phonon amplitude, which can be written as $\tilde{V} = d Q_\mathrm{H}^3 + e Q_\mathrm{H}^2 Q_\mathrm{G}^2 + f Q_\mathrm{H}^4 + g Q_\mathrm{G}^4$. We find the coupling coefficients $d$ to $g$ to be small compared to $c$. Substituting $V=V_\mathrm{min} + \tilde{V}$ into Eq.~(\ref{eq:eom}), we obtain
\begin{eqnarray}
\ddot{Q}_\mathrm{H} & + & \kappa_\mathrm{H}\dot{Q}_\mathrm{IR} + \Omega_\mathrm{H}^2Q_\mathrm{H} + c Q_\mathrm{G}^2 + \frac{\partial \tilde{V}}{\partial Q_\mathrm{H}} - Z_\mathrm{H} E = 0, \label{eq:eom_Higgs} \\
\ddot{Q}_\mathrm{G} & + & \kappa_\mathrm{G}\dot{Q}_\mathrm{G} + (\Omega_\mathrm{G}^2 + 2 c Q_\mathrm{H})Q_\mathrm{G} + \frac{\partial \tilde{V}}{\partial Q_\mathrm{G}}  =  0.
\label{eq:eom_Goldstone}
\end{eqnarray}
The nonlinear phonon coupling acts as a temporal modulation of the eigenfrequency of the Goldstone mode in Eq.~(\ref{eq:eom_Goldstone}), $\tilde{\Omega}_\mathrm{G}^2(t)=\Omega_\mathrm{G}^2+2cQ_\mathrm{H}(t)$. When $Q_\mathrm{H}$ oscillates at double the frequency of $\Omega_\mathrm{G}$, $Q_\mathrm{G}$ is parametrically amplified. Once the Goldstone mode is excited, $Q_\mathrm{G}^2$ reciprocally acts as a driving force for the Higgs mode in Eq.~(\ref{eq:eom_Higgs}).

Note that the phonon modes are detuned from the resonance condition $\Omega_\mathrm{G}=\Omega_\mathrm{H}/2$ by 0.4~THz. As we will see, this mismatch is compensated by the broadening of the phonon frequencies due to their natural linewidths and due to the temporal envelope that the IR-active Higgs mode obtains through the resonant driving by the terahertz pulse. In addition to the parametric downconversion described here, coupling of the Higgs mode to higher-frequency $A_1$ modes of the form $Q_\mathrm{H}^2Q_{A_1}$ may lead to a phonon upconversion according to sum-frequency ionic Raman scattering \cite{Juraschek2018}. This process is possible for coupling to the 7 and 9~THz $A_1$ modes in InMnO$_3$. As for other higher-order anharmonicities, this process would lead to quantitative changes in the amplitude of the Higgs mode, but not qualitatively influence the coupled Higgs-Goldstone dynamics described here. This coupling to other modes is implicitly taken into account in the phonon linewidths.

We model the electric field component of the terahertz pulse as $E(t) = E_0 \text{exp}(-t^2/(2(\tau/\sqrt{8\text{ln}2})^2)) \cos(\omega_0 t)$, where $E_0$ is the peak electric field, $\omega_0$ is the center frequency, and $\tau$ is the full width at half maximum duration of the pulse \cite{Juraschek2018}. In order to couple to the dipole moment of the Higgs mode, the electric field component has to be aligned with the hexagonal axis of the crystal. We choose experimentally feasible values for $E_0$ and $\tau$ for this frequency range \cite{Liu2017}. We list the used parameters in Table~\ref{tab:inmno3}. As the calculation of phonon linewidths is computationally challenging for larger unit cells (calculations for 2 or 4 atom unit cells have been performed \cite{Togo2015}, compared to the 30 atom unit cell in the present case), and no experimental data on phonon linewidths in InMnO$_3$ exists to our knowledge, we take $\kappa_\mathrm{H}$ and $\kappa_\mathrm{G}$ as free parameters of the model within possible experimental boundaries. Raman and IR studies on closely related hexagonal yttrium manganite (YMnO$_3$) show that phonon linewidths lie within 3-20\%{} of the phonon frequency, $\kappa_i \approx 0.03$ to $0.2 \times \Omega_i/(2\pi)$ \cite{Iliev1997}. Finally, we add a thermal noise $(2k_\mathrm{B}T\kappa_\mathrm{G/H}^2)^{1/2}\xi(t)$ to the equations to ensure that the initial amplitudes of the Higgs and Goldstone modes are nonzero at room temperature. Here, $\xi$ is a Gaussian random variable with variance one, $k_\mathrm{B}$ is the Boltzmann constant, and $T$ is the temperature \cite{Joubaud2007,Gitterman2013}.

We show representative dynamics of the Higgs and Goldstone modes after the optical excitation according to Eqs.~(\ref{eq:eom_Higgs}) and (\ref{eq:eom_Goldstone}) in Figs.~\ref{fig:phonondynamics}(a) and (b), where we have chosen linewidths of 10\% of the phonon frequencies. While the amplitude of the Higgs mode, $Q_\mathrm{H}$, reaches its maximum during the duration of the terahertz pulse in Fig.~\ref{fig:phonondynamics}(a), the amplitude of the Goldstone mode, $Q_\mathrm{G}$, only starts building up with a time delay of 2~ps due to the parametric amplification by the Higgs mode. The delayed build-up is also visible in the time evolution of the vibrational energy contained in each mode in Fig.~\ref{fig:phonondynamics}(b). The vibrational energy contained in the resonantly driven Higgs mode (scaling quadratically in both the eigenfrequency and amplitude) is roughly three times larger than that contained in the Goldstone mode. Note that the time delay of the build-up of the Goldstone mode amplitude depends on various factors, such as the relative frequencies of the phonon modes, their dampings, and the strength of their nonlinear coupling.

As the parametric excitation requires an initial nonzero value of the Goldstone mode, the outcome of Eqs.~(\ref{eq:eom_Higgs}) and (\ref{eq:eom_Goldstone}) depends stochastically on the thermal noise background. In order to estimate the magnitude of induced atomic motion, we therefore sample the phonon amplitudes $Q_\mathrm{H}$ and $Q_\mathrm{G}$ for a large number of linewidths $\kappa_\mathrm{H}$ and $\kappa_\mathrm{G}$ between 3 and 20\%{} of the respective eigenfrequencies. The low-frequency Higgs and Goldstone modes involve mainly motions of the indium and oxygen ions. We extract the root mean squares of the largest ionic displacements, $d_\mathrm{H/G}=\text{max}_{n} | \mathbf{d}_{\text{H/G},n}/\sqrt{2} |$, which we show in Figs.~\ref{fig:phonondynamics}(c) and (d). The resonant driving of the Higgs mode by the terahertz pulse yields displacements of the indium ions of up to 3~pm, which corresponds to 1.5\% of the interatomic distance. These displacements lie well below the Lindemann instability criterion that predicts melting of the crystal lattice when the displacements exceed $\sim$10\% of the interatomic distance \cite{Lindemann1910,sokolowski-tinten:2003}. Surprisingly, the parametric excitation of the Goldstone mode yields indium ion displacements of up to 3~pm, comparable to the direct excitation of the Higgs mode, even for moderate linewidths. The induced displacements depend strongly on the linewidth of the Higgs mode $\kappa_\mathrm{H}$ and show no dependence on the linewidth of the Goldstone mode $\kappa_\mathrm{G}$. The maximal displacements of the Goldstone mode in Fig.~\ref{fig:phonondynamics}(d) show the expected stochastic fluctuations due to the thermal noise. In contrast, Fig.~\ref{fig:phonondynamics}(c) for the Higgs mode shows no fluctuations, because its amplitude reaches its maximum right after the arrival of the terahertz pulse and before the build-up of the Goldstone mode amplitude due to parametric downconversion as seen in Fig.~\ref{fig:phonondynamics}(a). The narrowest linewidths yield the largest amplitudes, despite having less overlap of components at the resonance condition, $Q_\mathrm{G}=Q_\mathrm{H}/2$, suggesting that a major contribution to the overlap comes from the temporal envelope imposed on the Higgs mode due to the excitation by the terahertz pulse. 


\paragraph*{Conclusion.}

We have shown theoretically that resonant driving of the IR-active Higgs mode in InMnO$_3$ leads to a subsequent coherent energy transfer to the silent Goldstone mode through parametric downconversion. This process allows the coherent excitation of both Raman-active and silent phonon modes in contrast to conventional difference and sum-frequency ionic Raman scattering that only allow coupling of Raman-active to IR-active phonons. The energy conversion efficiency of the parametric excitation thereby outperforms ionic Raman scattering, in which only a fraction of the energy is transferred to other modes \cite{fechner:2016,juraschek:2017,Juraschek2018,Gu2018}, and is at par with predictions for the transfer of energy in the resonant phonon upconversion process sum-frequency ionic Raman scattering \cite{Juraschek2018,Melnikov2018,Juraschek2019}. 

The coupling between the Higgs and Goldstone modes derives purely from symmetry considerations, and we expect that the dynamics described here are a general feature of the hexagonal manganites. Furthermore, the mechanism of parametric downconversion requires only the existence of fully symmetric IR-active modes and is therefore relevant to all materials with broken inversion symmetry. The lattice dynamics are strongly dependent on the linewidth of the IR-active mode, but only weakly dependent on the linewidth of the silent mode. This dependence makes it simpler to predict the effect in other materials, as IR spectra are much more broadly available than hyper-Raman spectra. The efficiency of the mechanism is maximal when the phonon modes fulfill the resonance condition $Q_\mathrm{IR}/2=Q_2$. We suggest that techniques used in the field of ultrastrong light-matter coupling could be leveraged here by tuning the frequency of the IR-active mode into resonance with the second phonon mode in an optical cavity in order to increase the effect \cite{Flick2018,Juraschek2019_3}.

We point out that the parametric downconversion described here is fundamentally different from the parametric amplification observed in a recent experiment \cite{Cartella2017}, in which the temporal modulation of the frequency of the optical phonon mode results from a nonlinearity of the dielectric function in the phonon amplitude and no phonon-phonon coupling is involved. While it has been shown that parametric resonances of phonons may influence the superconducting critical temperature \cite{Knap2016}, the mechanism here is fundamentally different from the parametric amplification of light by an optically excited Higgs mode in superconductors discovered recently \cite{Buzzi2019}. We further note that coherent energy transfer to a nearly silent phonon mode has been observed in a high-harmonic phonon upconversion process \cite{Kozina2019}.

Finally, we conjecture that the generation of a large population of silent (hyper-Raman-active) phonons could possibly open a route towards stimulated hyper-Raman scattering in solids, which has before only been achieved in atomic vapors \cite{Vrehen1977,Cotter1977}.


\begin{acknowledgments}
We are grateful to Nicola Spaldin (ETH Zurich) and Hyeongrak Choi (MIT) for useful discussions and in addition to Nicola Spaldin for feedback on the manuscript. This work was supported by the DARPA DRINQS Program under award number D18AC00014 as well as the Swiss National Science Foundation under project ID 184259 and the ERC Synergy project HERO under grant agreement number 810451, by Harvard University, and by ETH Zurich. P. N. is a Moore Inventor Fellow supported by the Gordon and Betty Moore Foundation. Calculations were performed at the Swiss National Supercomputing Centre (CSCS) supported by the project IDs s624 and eth3 and at Harvard's Research Computing Facility.
\end{acknowledgments}



%

\end{document}